\documentclass{IOS-Book-ArticleX}     
\usepackage{graphicx}

\begin{document}
\begin{frontmatter}          
%
\title{Strong light-matter coupling: parametric interactions in a cavity and
free-space}
\runningtitle{Strong light-matter coupling ...}

%
\author[A]{\fnms{I. B.} \snm{Mekhov}
\thanks{Mekhov@yahoo.com, Present address: Inst. for Theor. Physics, Univ. of Innsbruck, Austria}},
\author[A]{\fnms{V. S.} \snm{Egorov}},
\author[A]{\fnms{V. N.} \snm{Lebedev}},
\author[A]{\fnms{P. V.} \snm{Moroshkin}},
\author[A]{\fnms{I. A.} \snm{Chekhonin}},
\author[B]{\fnms{S. N.} \snm{Bagayev}}
\runningauthor{I. B. Mekhov et al.}
\address[A]{St. Petersburg State University, V.A. Fock Institute of Physics, Russia}
\address[B]{Institute of Laser Physics, Siberian Branch of RAS, Novosibirsk, Russia}
\begin{abstract}
We consider parametric interactions of laser pulses in a coherent macroscopic
ensemble of resonant atoms, which are possible in the strong coupling regime
of light-matter interaction. The spectrum condensation (lasing at collective
vacuum Rabi sidebands) was studied in an active cavity configuration.
Parametric interactions under the strong light-matter coupling were proved
even in free space. In contrast to bichromatic beats in a cavity, they were
shown to appear due to interference between polaritonic wave packets of
different group velocities.
\end{abstract}


\end{frontmatter}


\section{Introduction}
We present a study of parametric interactions between laser pulses in a
coherent ensemble of two-level quantum objects (``atoms''). The main
attention is paid to the processes that can be obtained in the strong
coupling regime of light-matter interaction, which is achieved, when (i) the
high frequency of field-matter excitation exchanges exceeds the decoherence
rates, (ii) the external field is not strong enough to entirely determine the
evolution of a system: it is the reemission field that plays a key role and
provides the collective behaviour of atoms.

Recently, this regime has attracted attention in quantum optics with both
atoms and nanostructures such as quantum wells and dots (cavity QED, Dicke
effects, microcavity exciton-polariton parametric scattering, squeezing and
entanglement \cite{giacobino2}). In both fields, it is considered as one of
the key models for quantum information processing (QIP). In this context, the
works on single or small number of objects (photons, atoms, excitons) as
qubits should be mentioned \cite{raimond,vuckovic}.

During last years, there is a growing interest in the study of macroscopic
ensembles of quantum objects, which can be used as collective elements for
QIP protocols \cite{lukinRMP,DLCZ}. Quantum memories
\cite{lukinRMP,kupriyanov,polzikPRA04}, sources of single photons
\cite{kimblePRL04}, entanglement of ensembles \cite{duanPRL00} and new
sources of entangled fields based on optically dense atomic ensembles were
recently proposed. The advantages of such collective objects in contrast to
single-particle qubits were highlighted \cite{kuzmichPRL04}.

Investigations of the collective effects in the strong coupling regime may be
of especial importance, because in this regime, some purely quantum
properties of the phenomena do not decrease with increasing number of
particles \cite{rempekimblePRL91,carmichael}, which will be important for QIP
with both discrete and continuous variables. In Ref. \cite{giacobino2}, a
scheme for generation of twin polaritons (coupled light-matter excitations)
in semiconductor microcavities with quantum wells was developed. It was based
on the parametric scattering of microcavity exciton-polaritons in the strong
coupling regime. In this scheme, very strong quantum correlations between
polaritons were obtained inside a cavity. Nevertheless, since only one
polariton branch of the dispersion curve was used in the parametric process
proposed, quantum correlations in the outgoing light fields were essentially
reduced. In this report, we present a study of parametric processes both in a
cavity and free space, which arise due to collective energy exchange between
field and matter systems (i.e., beating between two polariton branches of the
dispersion curve) \cite{OS94,PRA03, PRA04}.

\section{Interactions in a cavity}

The most significant manifestation of the strong coupling regime consists in
the resonant density-dependent splitting of the dispersion curve into two
polariton branches. The splitting appears, when the collective coupling
constant (cooperative frequency) of a medium $\omega_c$ exceeds all rates of
incoherent relaxations $\gamma$:

\begin{equation}
\omega_c=\sqrt{2\pi d^2\omega_0 n /\hbar}=g\sqrt{N} \gg \gamma ,
\end{equation}
where $d$ and $\omega_0$ are the dipole moment and frequency of a transition,
$n$ and $N$ are the density and number of atoms, $g$ is the single-atom
coupling constant. Thus, in dense media, photons with equal wave vectors but
different frequencies exist.

If a spatial spectrum of a problem is fixed by a single-mode cavity, beating
between these frequencies corresponds to vacuum Rabi oscillations. A spectral
doublet, arising in this case, represents density-dependent vacuum Rabi
splitting of a cavity mode, which is proportional to $\omega_c$ (1). The
origin of this effect can be traced to linear interaction of light and
dipoles in a passive cavity \cite{carmichael}. The condition of weak field
(which cannot destroy polariton dispersion) is reduced to the statement that
the photon number is smaller then the number of atoms.

In this report, we present our results related to the strong light-matter
coupling in an active system: a multimode broadband laser with an intracavity
narrowband coherent resonant medium without population inversion. The
condition of the strong coupling (1) can be fulfilled with respect to that
macroscopic resonant absorber. A phenomenon of spectrum condensation was
analysed. It consists in the dramatic modification of a lasing spectrum under
transition from the weak- to strong-coupling regime of light-matter
interaction: instead of usual saturated absorption line, a bright narrow
doublet of generation appears.

\begin{figure}
\includegraphics{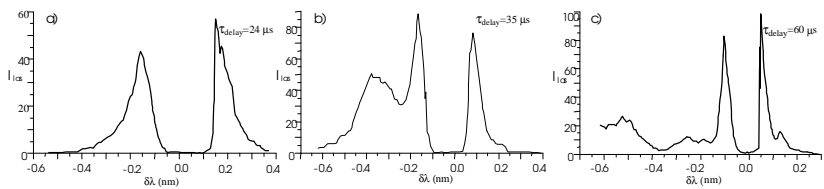}
\caption{\label{fig1}Self-splitting of a dye laser spectrum at different
moments of a Ne discharge afterglow, which corresponds to decreasing atomic
density in an intracavity cell, $\lambda=594.5$ nm.}
\end{figure}

Figure~1 presents dye laser spectra obtained with a neon discharge as an
intracavity absorber at different densities $n$ (up to $10^{13}$~cm$^{-3}$
corresponding to $\omega_c/2\pi$ up to 10~GHz) of neon atoms in the
metastable state, which was used as a lower state of a two-level system. It
was shown, that the density dependence of the doublet splitting is in
agreement with that of $\omega_c$ (1), which determines the vacuum Rabi
splitting. Moreover, the the existence of a density threshold, which is also
determined by Eq.~(1), was demonstrated. On the basis of the cooperative
parametric resonance model \cite{OS94}, this effect was treated as lasing at
collective vacuum Rabi sidebands due to parametric interactions between laser
modes.

\section{Strong coupling regime in free space}

The main difference of the free-space interaction consists in the fact that
the spatial spectrum of a problem is not fixed by a cavity. As a consequence,
under the linear propagation of a short pulse, no coherent density-dependent
spectral features (such as vacuum Rabi splitting in a cavity) can be
extracted from spectral measurements, except for the trivial appearance of an
incoherent absorption line. Nevertheless, in our work, we have shown that
such coherent features can be obtained under nonlinear parametric interaction
of laser pulses (the details of the theory can be found in Ref.~\cite{PRA04},
experimental results are presented in Ref.~\cite{PRA03}). In contrast to
cavity oscillations, they were shown to appear due to free-space optical
ringing, which does not originate from beating between waves of equal wave
vectors, but from the successive beats between polaritonic wave packets of
equal group velocities. Long coherent oscillations can be obtained due to
considerable reducing of the group velocity in a dense medium and due to its
essential variation over the broad spectrum of a short laser pulse. The
oscillation frequency at the initial stage of the ringing $\omega_D$ should
exceed the decoherence rates $\gamma$:

\begin{equation}
\omega_D=\omega_c^2 z/c \gg \gamma .
\end{equation}
This frequency increases with propagation distance $z$ and medium density
$n$. Thus, it is proportional to the number of interacting atoms, which
directly demonstrates the superradiant character of the field reemission by
the atomic ensemble.

\begin{figure}
\includegraphics{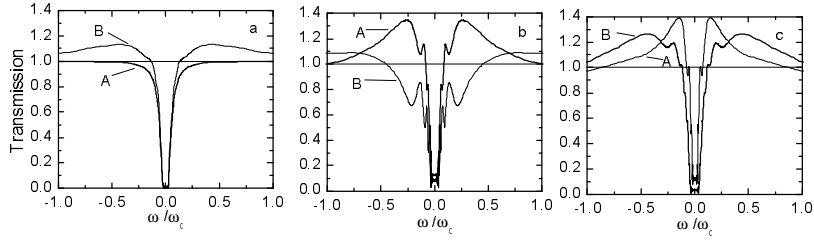}
\caption{\label{fig2}Transmission spectra of the probe field. (a) linear
transmission (A) and $\omega_c\tau_0 =0$ (B), $\omega_c z/c=1.0$; (b)
parametric amplification at $\omega_c\tau_0=-0.5$ (A) and dumping at
$\omega_c\tau_0 = 0.5$ (B) for $\omega_c z/c=1.0$; (c) parametric
amplification at $\omega_c\tau_0 = -0.5$ for $\omega_c z/c=0.5$ (A) and for
$\omega_c z/c=2.0$ (B); $\gamma/\omega_c=10^{-3}$, pump area $s_0=0.49\pi$.}
\end{figure}

We considered the nonlinear parametric interaction between two short laser
pulses propagating in a dense resonant atomic ensemble. Optical ringing was
shown to significantly affect the propagation and amplification of a probe
field under its interaction with a nearly copropagating pump. Depending on
the probe-pump time delay $\tau_0$ (the probe pulse precedes the pump for
$\tau_0 < 0$ ), the probe transmission spectra show either a specific doublet
or dip, which corresponds to parametric sideband amplification or dumping of
radiation [cf. Fig.~2(b)]. The widths of these features are greater than the
width of an incoherent absorbtion contour [shown in Fig.~2(a)], they are
determined by the density-dependent field-matter coupling constant and
increase during the propagation [cf. Figs.~2(b), (c)].

The spectral features in Fig.~2 can be explained as a result of parametric
forward scattering of a probe pulse on spatiotemporal modulations of the
population difference, appearing due to the optical ringing in a pump
\cite{PRA04}. The characteristic frequencies, which increase with the number
of interacting atoms, correspond to the frequency of optical ringing. The
condition of the weak fields, which do not destroy dispersion and collective
behaviour of atoms, in this case, is equivalent to the requirement of the
small input pulse area (so that $0\pi$ ringing should not be shadowed by
2$\pi$ solitons and Rabi flopping). Contrary to strong-field parametric
processes (e.g., due to transient Rabi flopping or stationary Mollow-Boyd
effect, which are determined by the pump amplitude), the density- and
coordinate-dependent spectra of the probe display the importance of
free-space collective oscillations and cannot be obtained in the framework of
a single-atom model.


\end{document}